# Experimental demonstration of drone-based quantum key distribution

Xiao-Hui Tian[1]*, Ran Yang[1]*, Hua-Ying Liu[1†], Pengfei Fan[1,2], Ji-Ning Zhang[1], Changsheng Gu[1], Mengwen Chen[1], Mingzhe Hu[1,2], Feng-Yu Lu[3], Zhen-Qiang Yin[3], Zhi-Jun Yin[2], Mo Yuan[2], Shuang Wang[3], Wei Chen[3], Yan-Xiao Gong[1†], Shi-Ning Zhu[1] and Zhenda Xie[1†]

*1. National Laboratory of Solid State Microstructures, School of Electronic Science and Engineering, School of Physics, College of Engineering and Applied Sciences, and Collaborative Innovation Center of Advanced Microstructures, Nanjing University, Nanjing 210093, China*

*2. Xin Lian Technology co., Ltd., Huzhou 313399, China*

*3. CAS Key Laboratory of Quantum Information, University of Science and Technology of China, Hefei 230026, China*

* These authors contributed equally to this work.

[†]corresponding authors：

liuhuaying@nju.edu.cn

gongyanxiao@nju.edu.cn

xiezhenda@nju.edu.cn

Quantum state transferring has been demonstrated using drones via entanglement distribution. Here we demonstrate the first drone-based quantum task, for quantum key distribution (QKD). Compact and polarization-maintaining acquisition, pointing, and tracking system and QKD modules are developed and loaded on a home-made octocopter, within takeoff weight of 30 kg. Real-time QKD is performed over 200 m distance with 8.48 kHz average secret key rate, using polarization-coded decoy-state BB84 protocol. With the capability of secret key distribution using a drone, wireless communication can be expected with enhanced security in the quantum approach, between mobile nodes towards a network.

Quantum key distribution (QKD) can generate and distribute random secret keys, to enhance the security of the communication technology in the quantum approach [1–12]. So far, QKD has been demonstrated using fiber or satellite [13–19]. They effectively work as the equivalents of the backbone network in the classical communication, and can cover key nodes over large distance up to hundreds or thousands of kilometers. However, the classical communication can reach end users in the form of wireless mobile communication, and its quantum resemble is missing up to date [20,21]. The difference for such mobile quantum communication is that a practical quantum task, like QKD, is coded in single photons, and cannot be copied and broadcasted in all directions from a fixed wireless base station. Flexible mobile nodes, like unmanned vehicles, are capable to establish wireless mobile quantum links for single photon signals, and reach the end users, either in standstill or motion, as required for a mobile quantum network construction [22–25].

Previously, entanglement distribution experiments have been demonstrated using drones [22,23]. These experiments only prove that high-fidelity photon state distribution is possible, however, a practical quantum task, like QKD, has not reported up to date using drone-based mobile quantum links. Even though much efforts have been taken for this goal [26–29], fundamental challenges remain to make the drone-based QKD possible. First, the QKD system need to be compact and stable enough to fit into the small drone, and work flawlessly in the continuous flying condition. Second, the quantum link between drone and ground or drones need to be built efficient, so that the overall link loss between QKD nodes is low enough to meet the fidelity requirement for a successful QKD. Moreover, the polarization of quantum states can be ever-changing with the beam pointing while the drone is in motion, so that the polarization control is necessary when the most popular polarization-coding is used for QKD.

Here we resolve all above challenges, and report the first quantum key distribution between a flying drone and a ground station. A 1.5 kg drone-based QKD transmitter module is developed, and paired with a ground-based QKD receiver module, to perform polarization-coded QKD using decoy-state Bennett-Brassard 1984 (BB84) protocol [30,31]. A 5 kg compact APT system is developed to be loaded on a small octocopter, which is capable for automatic acquisition to establish a quantum optical link within 30 seconds. The stable in motion polarization state transmission has been achieved, from the QKD module to the APT system, by sending two pairs of polarization states of horizontal/vertical and diagonal/anti-diagonal (noted as states $|H\rangle/|V\rangle$ and $|D\rangle/|A\rangle$) separately in two paths. In each path, a polarization-maintaining fiber (PMF) is used to carry each pair of orthogonal polarization states in the PMF polarization axes. With the above technologies that we developed, a low-loss and high-fidelity quantum optical link is established between the octocopter and the ground station that are separated 200 meters away, with an average link loss of around 9 dB. Real-time QKD is performed with assistance of classical wireless signal, and the average secret key rate exceeds 8 kHz during a 400 seconds test. With the capability of QKD using drones, secure communication in all mobile platforms in land, sea and air bases can be expected, thus this work marks an important step for the mobile quantum network.

The schematic of the experiment is shown in Fig. 1. We use the common downlink polarization-coded decoy-state BB84 protocol, via four-linear-polarization-state coding and two-decoy-state setting. A lightweight QKD transmitter module (QKD-TM) generates both the QKD photons and the classical optical pulse for temporal synchronization (SYNC). Both QKD photons and SYNC signal are distributed via a pair of APT systems, which track automatically between drone and ground station, and reach the QKD receiver module (QKD-RM) for decoding, detection and time synchronization, respectively. A wireless radio-frequency (RF) channel is also built between the drone and the ground station, for classical information transmission for real-time APT tracking and QKD post-processing towards key generation.

Before, only a simple spontaneous parametric down conversion source has been loaded on the drone in the entanglement distribution experiment [14,15]. To demonstrate the QKD using a drone, a full-function airborne QKD-TM is required in pair with a portable grand-based QKD-RM. For accomplishing decoy-state BB84 protocol that we used, such QKD-TM needs to be constructed with both optical and electronic components, for intensity manipulation, polarization coding, time synchronization and post processing, etc., which usually occupies a table-top size. Here we develop a compact QKD-TM, with its schematic shown in the red dash lines in Fig. 2. All the optical components, including laser diodes (LD), attenuators, beam splitters (BS) and polarization beam splitter (PBS) are micro-optics-based fiber components, so that the system polarization extinction ratio is only limited by the PMF pigtailing at about 20 dB. All the control electronics, including the laser diode pulse drivers for both QKD and the SYNC signal, the random number generation (RNG) circuit [32–37] and the post-processing circuit are all integrated in a single printed circuit board (PCB). As a result, the whole QKD-TM is packed in a small size of 179×179×60 $mm^3$ and weighs only 1.5 kg, including a 3D-printed box for protection against vibration, and temperature isolation. Inside the QKD-TM, the QKD signals are from eight 850 nm LDs, which are pulse-driven with 500 ps gates at 50 MHz repetition rate. Four each LDs are used for signal and decoy states generation, respectively. In each gate, one signal LD, one decoy LD or none is fired under the control of RNG, with probability of 50%, 25% and 25%, respectively. All the LDs are matched in wavelength and it is achieved by fine tuning the peak currents in the pulses. Since the output wavelength of the LDs are current dependent, different currents are applied to the LDs. The different driving currents also result in different output intensity, hence we apply individual attenuator for each LD, to achieve the relative intensities for signal states and decoy states, following the optimized values of μs = 0.73, μd = 0.20, respectively. Each fiber-coupled attenuator is fabricated with different misalignment for required attenuation between two c-lens collimators. All these TO-can-packaged LDs are glued to the same aluminum heat sink with active temperature control, with pins through the holes in the heat sink and soldered to their control electronics on the same PCB. Another SMF-pigtailed LD working at 808 nm, internally modulated by the same 50 MHz gate pulse train, is used for SYNC generation. And two PMF and one SMF are used for the QKD photon and SYNC signal output from the QKD-TM, respectively.

As shown in the blue and green squares in Fig. 2, a pair of compact APT systems, one installed on the drone and the other at the ground station, are used for the QKD experiment. Both APT systems feature identical optical aperture and mechanical design, including a 3-axis motorized gimbal stage for coarse tracking, and an optical assembly for beam collimation and fine tracking [38], as shown in Fig. 3 (a) and Fig. 3 (b). Initial acquisition is a fundamental challenge for establishing a drone-based quantum link. In previous experiments, such acquisition has to be initialized before the take-off, which limited the practical application of a quantum task. Here we develop a unique initial acquisition scheme, for the real-time coarse pointing angular resolution, so that the initial pointing can be automatically performed during the flight, as shown schematically in Fig. 3(d). Such initial acquisition is based on a deviation algorithm, combining the information from integrated inertial measurement unit (IMU) and real-time kinematic modules (RTK), to update the 3-dimension coordinates and coarse pointing angles in 10 Hz. As a result, the initial acquisition can be achieved either pre-flight or during flight on demand within 30 seconds, for a pointing accuracy of 8.8 mrad to enable optical tracking. Two-stage optical tracking is further performed to the beacon light BL1 and BL2, and feedback to the 3-axis gimbal stage and the fast steering mirror, respectively. We use a lightweight carbon fiber baseplate in each optical assembly. Together with the micro optics technology for polarization encoding/decoding units, the APT system only weighs 5 kg for both drone and ground station.

We develop a hybrid coding method to achieve high fidelity polarization state transmission between the QKD-TM and the APT system in the drone. This method uses two PMFs to transmit the polarization sets of $|H\rangle/|V\rangle$ and $|D\rangle/|A\rangle$ individually, where these two sets of the polarizations can be aligned to the fundamental polarization axes of the PMFs. In the QKD-TM, the $|H\rangle/|V\rangle$ and $|D\rangle/|A\rangle$ states, as required in the QKD protocol, are all coded in linear polarization, for PMF outputs. In the APT system, we use a polarization encoding unit (PEU) to combine QKD photons from these two PMFs, in the relative phases of the $|H\rangle$, $|V\rangle$, $|D\rangle$ and $|A\rangle$ polarization in a single free-space path, as shown in the orange dash line in Fig. 2 and Fig. 3 (c). This PEU also combines the SYNC signal with the QKD photons using a wavelength division multiplexer (WDM) into a single SMF. This SMF is taped down on the baseplate of the optical assembly, for a stable polarization state transmission to the collimation optics using an off-axis parabolic mirror (OAPM). In the free space optical link, we actively stable the roll angle of both the APT systems to gravity direction by gyroscopes, so their polarization frames maintain consistent during flight, and no additional compensation are needed during flight in local areas. As a result, the high-fidelity polarization state transmission can be achieved from the drone and directed to the quantum link, with the polarization contrast of both non-orthogonal quantum signals sets measured to be over 30:1 in the experiment.

In the experiment, we use a homemade octocopter to load the QKD-TM, the APT system, and the control electronics, with overall takeoff weight within 30 kg and flight duration exceeds 40 minutes, as shown in Fig. 3(e). During the QKD process, the octocopter is hovering at ~10 meters altitude, and the ground station is 200 meters away. We first test the quantum link during the flight, at a cloudy night with visibility of ~1 km, and wind speed up to ~25 km/h, as the typical weather condition in the season in Nanjing. The tracking errors in horizon/vertical directions on the drone and at the ground station are measured, with root mean square (RMS) values of 3.97/3.33 μm and 1.30/2.01 μm, respectively, as shown in Fig. 4(a). Such high tracking accuracy is enough for high collection efficiency from free space into SMF for the QKD photons, and the average link loss is measured to be ~9 dB. The total system loss is ~17.8 dB in average, including 8.8 dB excess loss from the optical components and the limited detection efficiency of the single photon detector.

At the ground station, the collected QKD photons and the SYNC signal are separated in the photons decoding unit (PDU) for key generation and time synchronization, respectively. The PDU also decode for the high-fidelity polarization state propagation from the APT system to the QKD-RM using two PMFs, in a similar way to the case on the drone. Inside the QKD-RM, the QKD photons are polarization projected and detected by single-photon detectors, following the timing generated by the SYNC signal. The post-processing is accomplished with assistance from the information through the RF channel, for generating the secret keys (details are shown in the supplementary material). The QKD can be performed at different weather conditions, either during night or daytime with illuminance under ~3000 lx. At night it can be limited from the condensation problem in wet weather on the optical parts in the APT system, while we can always perform the measurement during daytime with high success rate. Fig. 4 (b), (c) and (d) are results of three runs of our experiment, with the former performed at cloudy night and the latter two performed at daytime, respectively. The distribution is performed during measurement time of 400 s, 200 s and 200 s, and the average sifted key rates are displayed at every ten seconds, with values varying between 6.9 kHz and 23.4 kHz, 4.9 kHz and 25.9 kHz, 5.3 kHz and 23.2 kHz, respectively. We sample 10% of the raw data under both X and Z basis to estimate the quantum bit error rate (QBER), with values varying between 2.22% and 2.32%, 2.24% and 5.19%, 3.16% and 6.87%, respectively. The secret key rates are then calculated to be 8.48 kHz, 6.49 kHz and 5.33 kHz, respectively. It is worth mention that to suppress the background noise, we use both spectral and spatial filter and the time gating in this experiment. 3 nm narrowband spectral filters are used at the ground station, together with the SMF collection scheme, which effectively reduce photon leakage from the environment and the beacon light. In the post-processing process, single photons are recorded in the start-stop mode triggered by the SYNC signal, which removes the background noise counts outside the coincidence window. As a result, high signal to vacuum state ratio of over 50:1 and 30:1 are achieved at night and daytime respectively, which is important for our successful QKD demonstration.

In conclusion, we demonstrate the first quantum key distribution between a flying drone and a ground station, with maximum average secret key rate exceeding 8 kHz and 6 kHz during night and daytime, respectively. Compact QKD module and APT systems are developed, along with the scheme for stable-in-motion polarization state transmission technology, are the keys for this achievement. We use identical optical and mechanical design for both APT system in the drone and ground, to fit in the payload capability of both drone and a portable ground station. Such scheme can also be easily adapted into two mobile nodes, like two drones, for air-to-air QKD. Here the link distance is limited by the flying regulation of the field, so that the drone-ground station distance is 200 meters maximum, while our APT system is tested in free-space link with

sub-15 dB loss in free space up to 1.5 kilometer distance. Furthermore, as the photons are all coupled into single-mode fiber, multi-node mobile QKD can be expected using the multi-drone connection technology [23], for the increase of link distance and complexity of the connection topology. Here we focus on the local-area quantum network connection by the use of vertical-takeoff-and-landing-capable multi-rotor type of drone. With the fly-in-motion APT technology that we develop here, broad-area QKD can be expected using fixed-wing high-altitude unmanned aerial vehicle in the future, towards full-time all-location quantum network [25].

We thank Xiang Wang and Binbin Xu for helpful discussions. This work was supported by National Natural Science Foundation of China (62293523, 62293520, 62305156), National Key R&D Program of China (No.2019YFA0705000, No.2022YFA1205100), Leading-edge technology Program of Jiangsu Natural Science Foundation (No.BK20192001), the Excellent Research Program of Nanjing University (ZYJH002), Jiangsu Funding Program for Excellent Postdoctoral Talent.

## Figures

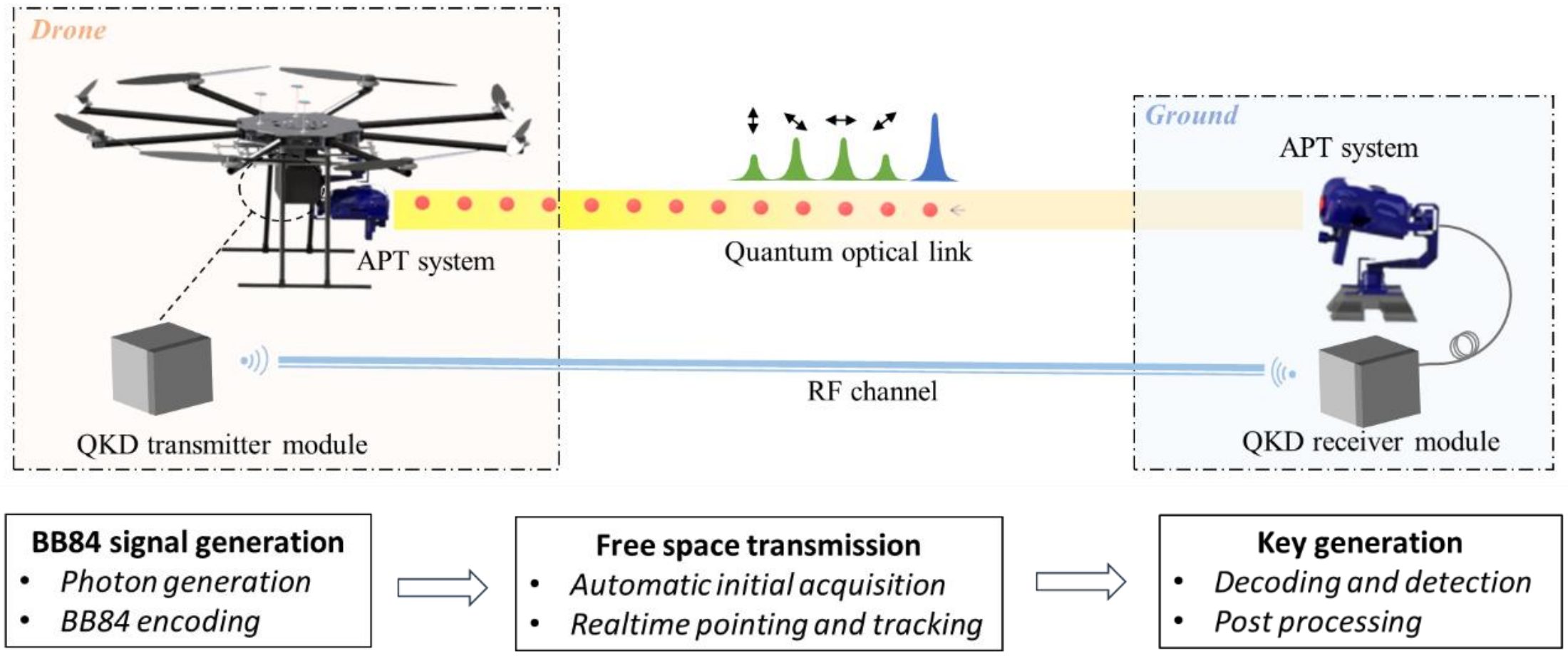

**Fig. 1. Schematic of drone-based QKD.** A compact QKD system using polarization-coded decoy-state BB84 protocol is developed. A robust quantum optical link is built between the drone and the ground station using acquisition, pointing, and tracking (APT) system for quantum signals (850 nm) and synchronization pulses (808 nm) transmission. Another radio-frequency (RF) channel is built for classical information transmission.

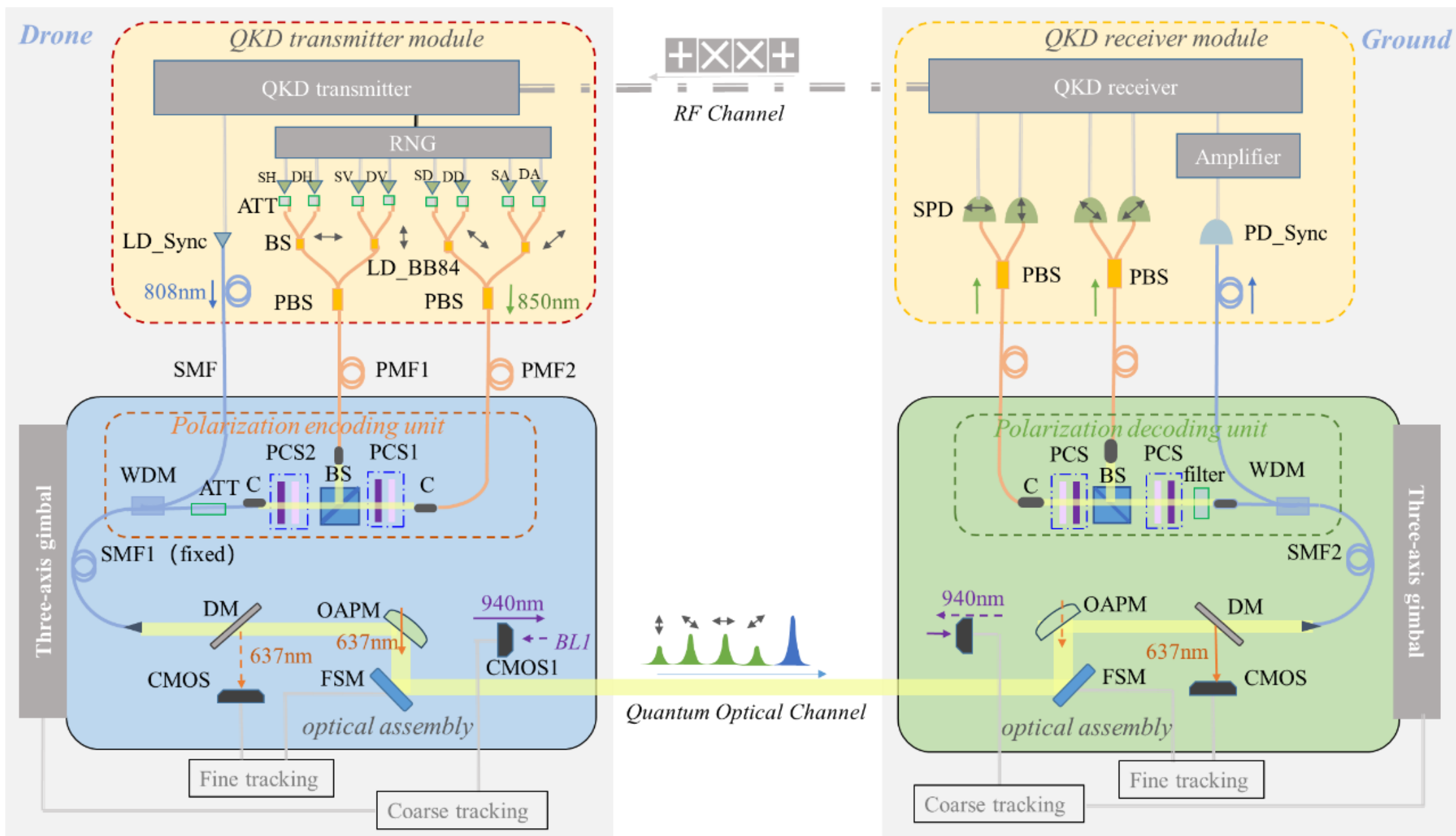

**Fig. 2. Details of drone-based QKD process.** The signal photons are generated in the QKD transmitter module and transmitted through two polarization-maintaining fiber to the APT system. Then the photons are encoded in the polarization encoding

unit. After distributing from the drone to the ground through quantum optical link, the photons are then decoded and detected on the ground in the reverse order. RNG: random number generator, LD_Sync: laser diode for time synchronization, LD_BB84: laser diode for decoy state and BB84 signal state generation, PBS: polarization maintain beam splitter, PMF: polarization-maintaining-fiber, C: collimator, PCS: polarization control system, BS: beam splitter, ATT: attenuator, PD_Sync: photodetector for time synchronization, WDM: wavelength division multiplexer, SMF: single mode fiber, DM: dichroic mirror, OAPM: off-axis parabolic mirror, CMOS: complementary metal oxide semiconductor, FSM: fast steering mirror, SPD: single-photon detector.

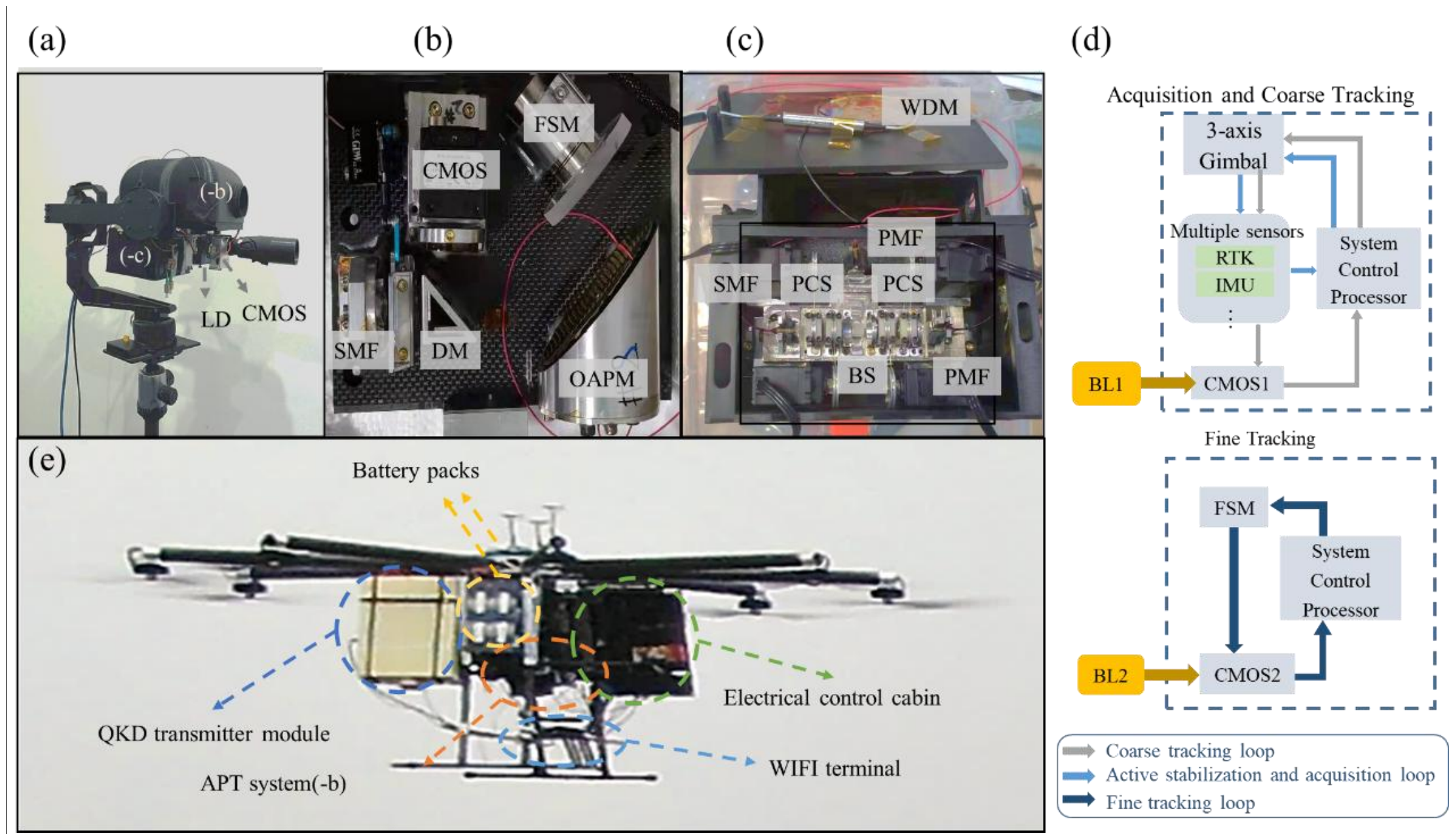

**Fig. 3. Pictures and diagram of the drone-based QKD system.** (a) The APT system at ground station. (b) The collimation and coupling optical circuit of the APT system. (c) The polarization encoding unit. (d) The schematic diagram of the APT system. (e) The flying octocopter in experiment. IMU: inertial measurement unit, RTK: real-time kinematic modules, BL: beacon light.

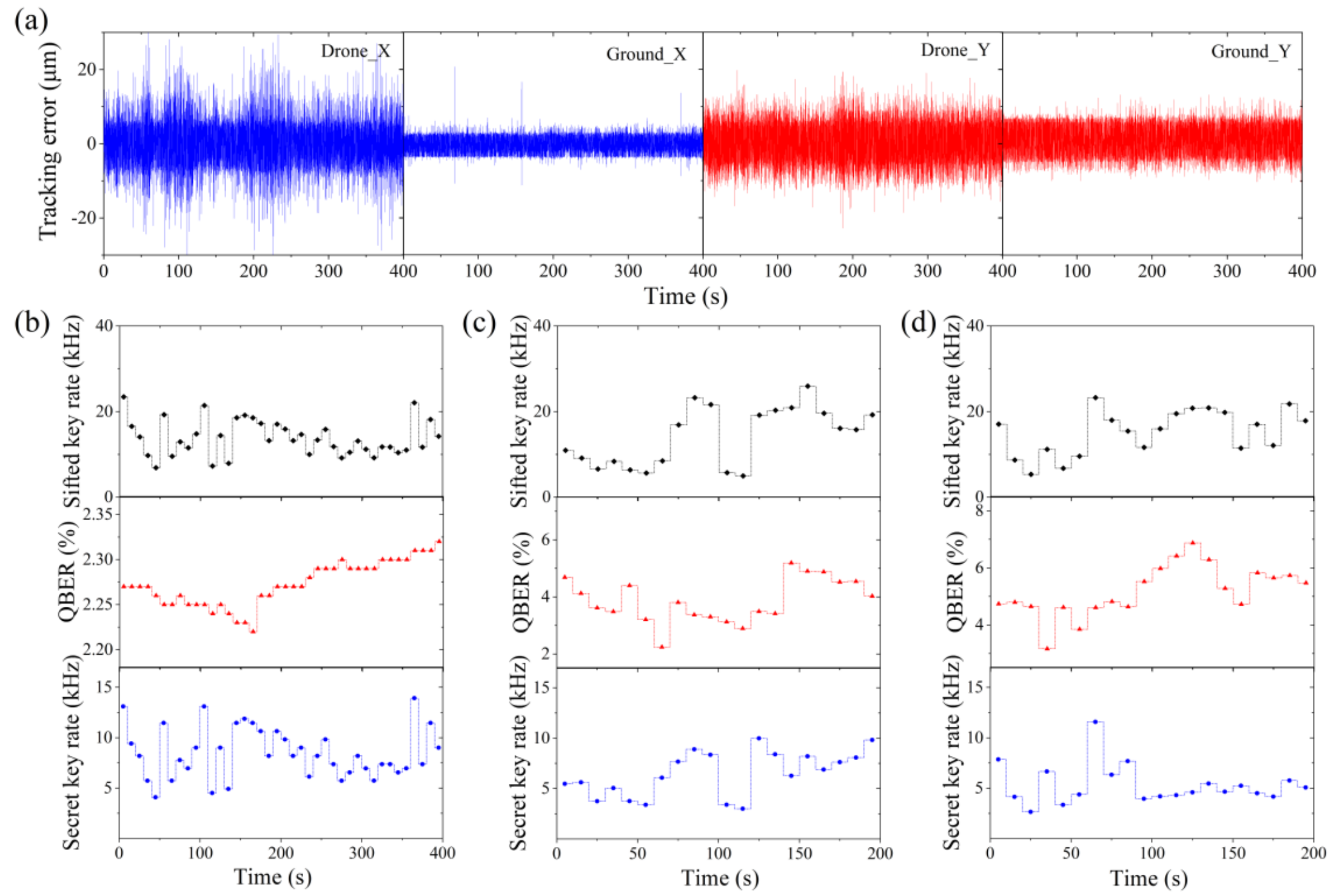

**Fig. 4. Experimental results of drone-based QKD.** (a) Tracking errors in both horizon/vertical directions are recorded by onboard data logging system on the drone and at ground station respectively. The results of quantum key distribution displayed during a time period of 400 s, 200 s and 200 s at night (b) and at daytime (c), (d), respectively, with the average sifted key rate, quantum bit error rate (QBER) and secret key rate displayed at every ten seconds.